# Misconceptions, Pragmatism, and Value Tensions: Evaluating Students' Understanding and Perception of Generative AI for Education

Aditya Johri
*Information Sciences and Technology*
George Mason University
Fairfax, USA
johri@gmu.edu

Ashish Hingle
*Information Sciences and Technology*
George Mason University
Fairfax, USA
ahingle2@gmu.edu

Johannes Schleiss
*Artificial Intelligence Lab*
Otto-von-Guericke-Universität Magdeburg
Magdeburg, Germany
johannes.schleiss@ovgu.de

*Abstract*—In this research paper we examine undergraduate students' use of and perceptions of generative AI (GenAI). Although the initial hype around ChatGPT has subsided, GenAI applications continue to make inroads across learning activities. Like any other emerging technology, there is a lack of consensus around using GenAI within higher education. Students are early adopters of the technology, utilizing it in atypical ways and forming a range of perceptions and aspirations about it. To understand where and how students are using these tools and how they view them, we present findings from an open-ended survey response study with undergraduate students pursuing information technology degrees. Students were asked to describe 1) their understanding of GenAI; 2) their use of GenAI; 3) their opinions on the benefits, downsides, and ethical issues pertaining to its use in education; and 4) how they envision GenAI could ideally help them with their education. Thirty-seven students provided responses ranging in length from 20 to 300 words for each question. Responses were iteratively coded by researchers to uncover patterns in the data and then categorized thematically. Findings reveal that students' definitions of GenAI differed substantially and included many misconceptions - some highlight it as a technique, an application, or a tool, while others described it as a type of AI. There was a wide variation in the use of GenAI by students, with two common uses being writing and coding. They identified the ability of GenAI to summarize information and its potential to personalize learning as an advantage. Students identified two primary ethical concerns with using GenAI: plagiarism and dependency, which means that students do not learn independently. They also cautioned that responses from GenAI applications are often untrustworthy and need verification. Overall, they appreciated that they could do things quickly with GenAI but were cautious as using the technology was not necessarily in their best long-term as it interfered with the learning process. In terms of aspirations for GenAI, students expressed both practical advantages and idealistic and improbable visions. They said it could serve as a tutor or coach and allow them to understand the material better. We discuss the implications of the findings for student learning and instruction.

Funding: U.S. NSF Awards 2319137, 1954556, USDA/NIFA Award 2021-67021-35329.

## I. INTRODUCTION

The use of generative artificial intelligence (GenAI) applications such as *ChatGPT*, *Gemini*, *Dall-E*, and *Midjourney* is finding increased use across academia and it is important to assess how these applications are starting to impact learning. As recent research shows, there is no consensus as yet on how these technologies should be used [1] and higher educational institutions are integrating GenAI in different ways across the institution and curriculum. Anecdotally, significant information exists about students' potential and actual use of GenAI, but there is a paucity of research studies that have examined this issue systematically [2].

Students are one of the primary users experimenting with GenAI, often driving the use of GenAI through early adoption, making it imperative to understand from their perspective how they use these applications and for what purposes. This is important given the hype and debate around the use of GenAI for plagiarism by students and the potential efforts to limit its use. By taking a restrictive stance, important educational opportunities and uses might be missed. Furthermore, providing students AI literacy and expertise, and skills such as prompt engineering [3], is essential for preparing them for the future workforce [4].

In this paper, we present a research study of the use of GenAI by undergraduate technology students. The study examined their understanding of GenAI and their aspirations for the use of GenAI for their education. It also examined their perceptions related to the benefits of the GenAI, and ethical issues with its use. We collected data through a series of open-ended questions and analyzed the responses thematically. We found that students have misconceptions about what GenAI is, even when they use it frequently, and largely see it as a tool for making things more efficient. They are also conflicted by its use as they see its benefits but realize it might impede learning. Ideally, they want GenAI to be a coach or tutor that can personalize the learning experience for them. We discuss the implications of these findings. This work contributes to research on the use of GenAI in education.

## II. RELATED WORK

### A. GenAI in Higher Education

Since the launch of ChatGPT in 2022, GenAI applications have become more accessible to a broad audience, and their capacity seems to increase at a rapid pace, ranging from performance improvements, fine-tuning to specific contexts or generation across modalities of image, text, speech, and video. At its core, GenAI models are trained on large datasets and often made available for inference as trained models. In the inference step, a GenAI model can generate an entirely new output that resembles the learned patterns based on a user input, often called a prompt. As the outputs are generated as a statistical inference based on the training, each output is unique, and it is not possible to reconstruct.

GenAI applications can be used in different functions within higher education utilizing its different capabilities. Widely referred use cases include the generation of content such as reports, assessments, or lesson plans, summarizing and supporting research, personalized learning experiences through the adaption of materials or individualized feedback, supporting administrative tasks and enhancing accessibility for students with disabilities, such as generating audio descriptions, transcribing lectures, or providing real-time captioning [5]. Challenges and concerns include discussion about academic integrity in assessments, impact on the learning process, for example, through cognitive offloading [6], clear ethical guidelines, and unknown accuracy and reliability of the models [5]. Overall, there is still uncertainty on the impact these tools have on the learning process and experience as well as the roles of the educators [5].

### B. Use of GenAI among students and faculty

Acknowledging that GenAI models will only increase their capabilities over time, it is important to co-construct educational experiences with GenAI applications and all stakeholders in mind [7]. To do so, it is important to understand the perceptions of involved stakeholders in the process. This also aligns with Biggs' 3P model (presage, process, product), highlighting student perceptions' influence on learning outcomes and approaches [8].

Previous research on educators' and students' perceptions and concerns has focused on understanding familiarity with GenAI tools, frequency, and type of use, as well as contrasting that in different domains. As the technological adoption and capabilities of tools increase these can always only be snapshots at a certain time.

A study among students in Hong Kong demonstrated that students are overall familiar with GenAI tools and linked the level of familiarity to their frequency of use and knowledge about GenAI [9]. Similarly, Singh et al. [10] studied the use and perception of ChatGPT by Computer Science students in the UK in their learning activities. Their findings indicate that students, while being familiar with the tools, do not regularly use them and are skeptical of positive influences on their learning. Moreover, the data indicates that students lack a deeper understanding of the functionality and proper use of GenAI tools.

Different factors influence the perception and use of GenAI tools. [11] investigated the correlation of Theory of Planned Behaviour (TPB) model factors of attitude, subjective norms, and perceived behavioral control and the perceived benefits, strengths, weaknesses, and risks of GenAI tools among students and lecturers. The study indicates that the attitudes, subjective norms, and perceived behavioral control positively correlate with the perceived strengths and benefits of GenAI tools. At the same time, the study finds that the perception of weaknesses and risks of GenAI applications vary between students and educators. Similarly, [12] showed that cultural dimensions and cultural expectations influence respondents' perceptions regarding the benefits and concerns associated with GenAI tools, especially related to academic dishonesty and the necessity for ethical guidelines. Another relevant factor to consider is the generational differences in the experiences, perceptions, knowledge, concerns, and intentions of using GenAI tools [13]. Chan and Lee [13] find that the younger generation (Gen Z) was more optimistic, while teachers from the older generations (Gen X and Gen Y) recognized potential benefits but also highlighted concerns regarding over-reliance on tools, as well as ethical and pedagogical implications.

### C. Technology Adoption

Finally, beyond the specific prior work related to GenAI, there is also literature on technology adoption more generally that is of relevance here. This literature shows that there is always ambivalence around any new technology, and a range of perspectives are to be found [14]. For many potential users, the pragmatic advantages take precedence over other considerations, but for others, the ambivalence stays and often turns into value tensions around using technology [15]. Another interesting element of adoption is users' aspirations based on their understanding of what the technology is or is not capable of. Often, the aspirations are misinformed due to misconceptions users have about how the technology works. This can result from media portrayals of technology or lack of knowledge of technology or both [16].

One recent noteworthy example of this is the idea of 'self-driving cars.' Advances in computer vision and machine learning have resulted in big strides towards a car's capability to drive without human intervention, but the capability to drive without any human input still remains far-fetched [17]. Yet, for many early adopters, the current advances and expectations have been enough of a motivator to adopt cars with self-driving capability, whereas, for many others, there is an ambivalence around it. Since many cars with this functionality are electric vehicles, there is also a value tension around the promise of eco-friendliness given the climate-damaging reality of building new cars and batteries.

Overall, it is clear from the literature that various viewpoints and aspirations are common as new technology is introduced and adopted. This is especially the case with applications or systems that are or seem to be transformative, as is the case with GenAI. Therefore, studying and understanding what different user groups think and expect from the technology is even more important. Hence, in this research study, we investigated students' views on their use and adoption of GenAI for education better to understand their motivation, concerns, and aspirations.

## III. RESEARCH STUDY

Consistent with human subjects research protocols, this study was approved by the Institutional Review Board, and data were collected only from participants who consented to the study. Each student submitted a consent form at the start of the semester, and data for any student who did not consent was deleted by a member of the research team who was not associated with teaching or grading the course. This member also anonymized all data and assigned unique numbers to each participant.

### A. Study Site and Participants

Data was collected from students in an undergraduate course on technology and society. Students in the course were enrolled in an information technology degree with different specific majors such as networking, web design, and cybersecurity. Students were either juniors (3rd year) or seniors (4th year). The institution where the data was collected is regularly ranked as the most diverse university in that state and one of the most diverse across all of the United States. The class composition reflected that many first-generation students and first-generation immigrant students represent a range of nationalities and cultures. The institution does not make precise demographic data available. A majority of students in the class are transfer students from local community colleges. Forty-four students were enrolled in the class, and after removing those who did not consent and missing responses, responses from thirty-seven students were included in the analysis.

### B. Survey Design

The data was collected using a short survey with open-ended questions. Since the use of GenAI is a relatively new phenomenon and studies of student perceptions are limited, an empirical decision was made to collect more open-ended data rather than responses to specific survey items. Open-ended responses also allowed the collection of a larger range of views than those traditionally possible through closed or Likert-style questions.

The following questions were given in the survey:
1) What, according to you, is Generative AI (GenAI)? What GenAI applications have you used?
2) Discuss your use of GenAI, providing specific examples of how you have used it. Specify if you use it regularly or you have only tried to use it.
3) What do you think are the potential benefits of using GenAI for education? What are some downsides?
4) What ethical issues do you think exist or do you foresee with the use of GenAI applications for education?
5) In an ideal world, how GenAI can help with you with your education?

### C. Data Collection

The survey was administered as an in-class activity as part of the course and was a preliminary exercise before an assignment where they used GenAI was assigned to them. The survey was administered online through the course's learning management system.

### D. Data Analysis

Data analysis was done iteratively [18]. First, two authors read through all the responses and discussed the overall themes present in the responses. Second, the first author coded the data based on broad categories and discussed these with the second author. Third, the categories were refined, and both authors conducted further thematic analysis. As a final step, the third author compared the coding and ensured validity.

## IV. FINDINGS

In this section, we report findings from the survey data spread across five major categories: students' 1) understanding, 2) their use, 3) their perception of benefits and problems, 4) their view on ethical concerns, and 5) their aspirations. Tables I and II depict thematic categories identified across the data. The quotes and excerpts presented have been edited minimally for grammar and clarity, with specific attention not to alter intent or meaning.

### A. Students' Understanding of GenAI

Students expressed a range of understandings about what GenAI was and how it works, ranging from relatively good to quite naive or misconceptions. For many students, GenAI was largely the conversational aspect, such as applications like ChatGPT or Bard/Gemini. Some students also mentioned applications such as MidJourney or Dall-e or using GenAI as part of other software such as Adobe Photoshop.

*"Generative AI is the collecting of information from a large source to generate a response to the user's input. I have used ChatGPT before in the past."* [S37]

*"GenAI is trained on data sets and can figure out patterns to create content. It is able to generate responses, images and other forms of information like ChatGPT which is an application I've used before."* [S38]

*"Simply put, GenAI are applications that provide with a thought based on research (scouring the internet based on the input you give). Some that I have used are ChatGPT and Bard."* [S18]

On the expert side, students gave quite extensive explanations as well:

*"Generative AI is a system that synthesizes information and creates a complex algorithm based on the information given and how it was trained. An example of generative AI is something like ChatGPT or in my case I have used the generative AI from Adobe Photoshop to restore old and damage photos. Generative AI itself is a learning based on information and present form of medium often times choosing the best results or most common value or image. I have also used it in video in terms of editing to give me some inspiration towards transitions or composition."* [S17]

TABLE I
QUESTIONS, RESPONSE CATEGORIES, AND EXAMPLES ON GENAI DEFINITIONS, BENEFITS, AND DOWNSIDES IN EDUCATION

| Response Categories | Example Quotes |
| --- | --- |
| **What is GenAI?** | |
| Naïve/Folk | "Generative AI is the process of giving AI a prompt and it spitting out an image of what you are describing or at least its best ability." [S33] |
| Misconception or Partially Sound | "[GenAI] is artificial intelligence capable of generating data or any other information you need such as educational or general information." [S34] |
| Accurate | "GenAI is a program that learns from pre-made references, such as text or images, and uses what it is taught to create something new, usually based on whatever was used to teach it." [S29] |
| **How have you used GenAI?** | |
| Basic | "I've used it in emails and text messaging. When I'm typing sometimes a few words will pop up for me to use if I choose to use." [S16] |
| Intermediate | "I use ChatGPT for my work, I am the social media manager and it helps generate captions and popular hashtags to make certain posts trend, I use it regularly." [S03] |
| Advanced | "At my workplace, I am in charge of the IT department. I use ChatGPT4.0 to help create policies both for IT & HR, utilize it to speed up any programs, query it for questions that may not have straight forward answers, to do research on topics I know little about, and to analyze long technical documents into shorter condensed guides for employees to use." [S21] |
| Consumption only | "The time I used the GenAI was out of curiosity to generate random photos using random prompts that popped into my head. It was less about using it and more about seeing what GenAI could do." [S14] |
| **What are the potential benefits of GenAI for education?** | |
| Assist Educators | "I also think that it would bring convenience to educators as they can use it as a resource and understand how to teach their material to a diverse group of people." [S03] |
| Personalization to the Individual | "I think the potential benefits of GenAI for education is it is a completely tailored to the user. We all know that every student not only learns differently, but at different speeds and through different means as well." [S08] |
| Fast Access to Information | "I think that the biggest benefit that generative AI could do for education would be to help people get information quickly without having to search everywhere for it." [S40] |
| Make up for Gaps in Student Learning | "GenAI could help students outside the classroom with subjects they may have not fully absorbed during lectures, or maybe they didn't have time to fully review them." [S24] |
| Support different learning styles | "In ChatGPT I find it useful that AI is able to explain topics to me like I am a beginner so I can learn the fundamentals of something such as a snippet of code, in a consolidated manner as opposed to having to piece together my own research from several sources." [S08] |
| **What are potential downsides of using GenAI in education?** | |
| Lack of Accuracy | "Downsides include the AI being completely wrong, I've used some tools and the information was either outdated or just incorrect." [S03] |
| Students Not Actually Learning | "However, when it comes to students using GenAI, it can reduce the effort student put into problem solving, critical thinking, and creativity because they might potentially rely heavily on GenAI to do the heavy lifting for them." [S20] |
| Plagiarism and Credibility | "Plagiarism is also another downside as AI does not typically source where they got their information from." [S37] |
| Promotes Apathy and Laziness | "I think the main thing is students will take advantage of it and use it to get work done fast and in a lazy manner." [S22] |

*B. Students' Use of GenAI*

There was a wide variation in how students used GenAI. Some of them had tried it briefly, while others used it regularly. In addition to their school work, students reported using GenAI as part of their work (many students in the program work part-time or full-time).

"I have used GenAI like ChatGPT to help me with some homework like you can ask ChatGPT anything it will give you an answer with a detailed explanation. It also can help you research right now, like find a hotel or plane ticket." [S44]

"I have used it to check for grammar mistakes when writing important documents such as resumes. However, I do not use it regularly because I believe that relying too much on AI can impact our productivity and skills." [S34]

Students also reported using it extensively to learn:

"A way I have used ChatGPT were to create a basis of a website when I was beginning to learn front end development. It is still regularly being used in order to correct mistakes or teach me how to implement different libraries and basic things I forget how to do like imports. I have also used ChatGPT to help me outline different ideas, come up with names, and also generate images when DALL-E was first introduced to OpenAI. Siri was one that I used a few times when I was first introduced to Apple products for basic commands such as answering day to day questions (about weather, an obscure fact that I was trying to find, etc.) and to navigate through applications. I have come across some tools that were able to transcribe voice messages and turn them to text, although I am unsure if it counts as GenAI." [S42]

For many students, the capability of GenAI, especially ChatGPT, that they relied on most was to summarize large bodies of text:

"For some of my classes that share case studies with 10+ pages of information, I simply asked a GenAI application to give me a summary of the case and include only the most important parts. Then after getting a good understanding of the case, I am able to skim over the 10 pages or whatever with a better understanding of the main takeaways. It gives me all the main points so that when I review the actual class material, its more familiar to me and I have a better understanding." [S38]

Many students expressed a positive sentiment about their use of GenAI and stated that they would like to see higher integration of the technology within education:

"Personally, I do a lot of work with AI, and I am all for it. I have most notably been using AI in building applications. The most recent of which is an automated weather notification system that utilizes ChatGPT, and it has helped tremendously. By using the OpenAI API, I was able to capture a bunch of weather data every day and filter that straight into the OpenAI generative AI API. I would then prompt ChatGPT to format that raw weather data into a more clean and readable format and send me that output, which I can then send to users that want to know that information without ever having to check the weather, and it is also notably way faster. I enjoy using these technologies, and I hope they keep getting smarter." [S36]

Overall, students were using GenAI for a range of activities, and even if the use was not deep in terms of what they were doing for everyone, they seemed to be finding use across a range of activities:

"I have used ChatGPT. I often use it for brainstorming, that is I will ask it to give me certain ideas on topics. I also like to use it to summarize my work, for example when editing my resume, I wanted to shorten a job description, so I asked ChatGPT to do it for me. I have also used it to debug code, it can help point out where your code has errors. I would say I use it pretty regularly. For a hackathon, I created a language-learning website that used a language chatbot made by one of my teammates using generative AI." [S27]

C. *Students' Perception of Benefits and Problems with GenAI Use*

Students provided a set of diverse responses for how GenAI can benefit education. For many, the biggest advantage was easily getting an answer or response to a question or piece of information. They also referred to the potential capability of GenAI for personalizing information for a user or learner. This was an offshoot possibly of how they were currently interacting with the technology through a conversational interface such as ChatGPT.

"GenAI provides an easy way for someone to get the answers that they're looking for. However, it may not always be accurate and originality and self-learning is minimized." [S15]

"GenAI can change how we learn by making it more personalized. It looks at how each person learns and gives them lessons that fit them best. With GenAI, we can learn anytime, anywhere, and even play with simulations and experiments to understand things better. But we need to be careful about things like unfairness, and privacy and make sure the lessons are accurate. Working together, teachers and tech experts can make sure GenAI helps us learn better and fairly." [S19]

Students were also aware of the shortcomings of GenAI and the potential problems it could create. For instance, they stated that they knew not all responses were accurate and that the more they used the system, there was a possibility that the less they would learn a topic for themselves.

"Like how I use ChatGPT to explain concepts, check my grammar, and fact check, it really is very efficient and accessible. I definitely prefer having one clear explanation to a question instead of just searching it up on a search engine and have multiple sources telling you different things which then you need to do further research to make sure it's credible. I know CHAT isn't 100 percent reliable too, but for simple concepts that have been around forever, I trust it as I know it takes information from millions of websites and databases to form an answer. I feel like it does the work for me, as I am able to skip that step I used to have to do. Some downsides are that students can become too reliant on it, use it to cheat, plagiarize, not do any of their own work. It can allow students to not have to critically think which is dangerous as they won't actually be learning anything. It can also be incorrect and will give students misinformation." [S32]

For many students, integrating GenAI with their formal education was a benefit as it would allow institutions to scale up teaching. Many students referenced the high student-to-teacher ratio and the lack of personal attention for students or immediate response that held back their learning.

"Using GenAI can have many benefits when it comes to using it in education. It can make the grading process for teacher faster and easier, help students understand the topics they have a good understanding in, and so on. However, when it comes to students using GenAI, it can reduce the effort student put into problem solving, critical thinking, and creativity because they might potentially rely heavily on GenAI to do the heavy lifting for them." [S20]

"Potential benefits could be provided students a more personalized and hands-on education opportunity that a 1:50 teacher:student ratio may not be able to provide in the classroom, but with a GenAI assistant, students may be able to get a lot more out of their everyday learning. GenAI should be used to augment and help educate students but needs to be kept in mind that it should not fully replace teachers as interpersonal communication is a key skill. Downsides include students using the 'get out of jail free' card that GenAI may provide them in completing their assignments for them, completely undermining education attempts as if students won't even read or understand why they are doing their assignments you will have a lot of people who are not very knowledgeable over what they supposedly 'studied' if they rely heavily on GenAI." [S21]

TABLE II
QUESTIONS, RESPONSE CATEGORIES, AND EXAMPLES ON ETHICAL ISSUES AND IDEAL USES

| Response Categories | Example Quotes |
|---|---|
| **What ethical issues do you see or foresee for GenAI in education?** | |
| Student Privacy | "It may increase the risk of student information being exploited and compromise the privacy of students." [S22] |
| Copyright/ Intellectual Property | "I think the biggest ethical issues on the education for GenAI is Copyright and legal exposure." [S44] |
| Bias | "A lot of the information and responses we get from AI could also be biased because they are taken from different parts of the internet." [S23] |
| Plagiarism and Homogeneous Thinking | "I foresee ethical issues of using GenAI applications for education as producing same materials/results, or plagiarism among students using the same tool for their assignments etc." [S26] |
| Complacency | "Also the information posted by GenAI could be off, but people wouldn't know because they would be so reliant on it." [S38] |
| **In an ideal world, how can GenAI help you with education?** | |
| Free to Use as Needed | "In an ideal world I would be free to use AI to help me complete the tasks I need to do as long as I know how to use it and no one will blame me as long as I complete my job successfully and without error." [S44] |
| Efficiency for Task Completion | "GenAI has aided me in providing ideas and insight into the many ways I could take on a task..." [S15] |
| As a Tutor or Coach | "In an ideal world, GenAI can be my personal tutor. I can ask it to explain things in different ways until I understand." [S27] |
| Brainstorming Ideas | "It can become a tool to start out generate ideas and like another person in the room brainstorms potential topics in a case by case." [S17] |
| Information Summarization | "It also helps explain things in a simpler manner which in turn helps me retain information better. It is able to explain things step by step which complements my learning style. " [S23] |
| Generic Responses | "It is also great for helping students with writers block just to get an idea of how they want to structure their paper or key points that they didn't think about implanting." [S37] |
| Interest-based Learning | "I would love to be able to expand my creativity during assignments and have my work come to life right in front of me. I think the benefits of how it allows art and content generating would be great for me." [S22] |
| Scale-up Classes | "GenAI can help me by bridging the gaps from professor to myself. Say I have questions or not a clear understanding of topics, GenAI could be the 'middle-man' and give me a much more personalized education experience than a professor with 100s of students." [S21] |
| Professional Development and Professionally Informed | "As an IT student, I could get images or videos by the professors of the current IT industry's situations or news, or coding videos created by GenAI to help better understand coding." [S26] |
| Just-in-time Learning Support | "It can be like a second teacher. Instead of trying to get a hold of the teacher anytime I don't understand something, this is like a simpler way of learning..." [S18] |

### D. Students' Views on Ethical Concerns with GenAI

Students brought up a range of ethical concerns with the use of GenAI, ranging from plagiarism to copyright and intellectual property rights. They also expressed concern with how their instructors and institutions treated the rise of GenAI. They expressed the lack of consensus made weighing the benefits of the technology with the potential risk of being "caught" difficult for them as students.

"The ethical issues, as I previously mentioned, primarily lie in the student's use of GenAI to circumvent doing what they perceive to be trivial work. There is also an ethical quandary over how educators should use GenAI. On the one hand, some ban it's use outright. Others have no opposition, and there is no clear consensus on what the best way forward is for educational institutes. My belief is that GenAI is a tool like any other, and it should be used as such. Not shied away from, but instead instructed on how to use. How to form prompts, and how to check it's work. Anything less, and the institution is cheating it's students. That would not prevent students from using it, and the school will have to expend excess money and effort combating a problem that doesn't have to be." [S10]

"The main ethical issues that I foresee with the use of GenAI is that of cheating. GenAI is quick and convenient, making it a tool ripe for picking. A student can simply type in a prompt and use the results of that prompt for submissions. This essentially allows them to cheat the school system and not learn the material." [S14]

"Using GenAI in education can lead to problems like unfairness, privacy worries, making sure the information is right, and students relying too much on technology. Also, by using GenAI applications, the student won't learn anything if they just those apps for everything. Because then they become so dependent on it that they forget how to respond to things in their way and are not able to think that much." [S19]

"The main ethical issue I think genAI creates is copyright and hindering learning. The data used to create output usually

*not freely given and an argument can be made that the student is using unauthorized assistance which can prevent actually learning of a topic if the student is solely using genAI to finish assignments." [S28]*

### E. Students' Aspirations of GenAI

Students expressed a range of aspirations related to GenAI regarding their education. Many of these were quite broad and idealistic in terms of the functions they expected GenAI to fulfill. From helping them tutor and coach to assisting with professional development, students also expected the technology to be able to make judgments about what they would find interesting.

*"It can help me by informing me what I should be learning, keeping me aware of current issues, and by directly training me. This training can be in the form of interview question preparation, revision of academic papers with explanations, or brainstorming topics to discuss in classes. There are an endless variety of ways to use GenAI to improve your learning, and the limit is an individual's creativity." [S10]*

*"In an ideal world, GenAI can help me with providing accessible and educational materials, such as it can determine which sources is good and eliminate the bad sources so that the information for my assignment would be more accurate and professional. It could help me with offering real feedback and support when I need it instantly. Additionally, it could provide information that is align with my personal interest, goals, and most importantly up-to-date information." [S39]*

Some students, although they had high expectations, had actually tried out the application and had a much better assessment of its capabilities and what it could actually help them with.

*"In an ideal world, GenAI can help assist me learn with about topics without just explicitly giving me the answer. So far this has been true in many cases. Just the other day I was trying to print two lists at the same time on the same line in python. I had created a very inefficient way of doing such task that required more work and was more confusing. I asked GPT if there was a better way to complete such a task and I was informed of this new zip() function and was able to complete the task in one line. I have never heard of this function and likely would not of found out about it. I know this is just a small example but it really shows how AI can be used not just to give answers, but explore new ideas and help you build off an existing foundation of knowledge along with its many other uses that I discusses earlier like its help in development. I can learn from AI by asking questions that I am confused about to have a better grasp of such concepts that need further explaining." (S36)*

## V. DISCUSSION AND IMPLICATIONS

The use of Generative AI in education is in flux, given the dynamic nature of the technology and the lack of consensus on its application for teaching and learning. Consequently, students have a range of views on the viability of using GenAI for education. It is important, though, to understand students' perspectives so that as GenAI is adopted, it can be done in a manner that considers student feedback. A better knowledge of student perspectives can also help motivate faculty who are hesitant to integrate it to learn about it. Finally, given the disproportionate focus on the use of GenAI for plagiarism and the resulting academic integrity violations, this and related study can help shift the focus on some positive aspects of GenAI use.

**Misconceptions**: The first finding that is both interesting and troubling is the extent of misconceptions students have about GenAI in terms of what it is and how it works. Many students equated it with being like a search engine or something that summarizes information. Very few identified the predictive and machine learning aspects of GenAI. They also confused GenAI with other ML operations, such as translating text from audio. Overall, most students expressed an interest in using GenAI, and several reported using it regularly.

**Pragmatism**: For many students, using GenAI was a productivity tool that made doing their work more efficient. They could find information easily and even use it to summarize content they thought was too long. Finally, the availability of GenAI 24/7 was something students commented on and expressed this was important given that they could not always access their teachers or teaching assistants.

**Value Tensions**: Most students also stated that they understood that even though GenAI made some aspects of education much easier, this came at a cost – they learned less. This balance between using it productively and still being able to learn new things created a value tension within students. This tension was further exacerbated as the guidelines for using GenAI and those for plagiarizing are unclear and a moving target. For many students though, this was a cause for concern as they saw the use of GenAI as becoming the norm in the workforce and they wanted to be prepared. Others were not convinced that using GenAI helped them learn anything in the short term and hindered long-term learning. Their viewpoints got further muddled as some faculty allowed and even encouraged using GenAI while others prohibited it. The value tensions also existed as the inability or lack of learning due to the use of GenAI and plagiarism were identified by students as ethical concerns due to the technology. Some students also brought up copyright and intellectual property theft issues, given how the data was trained. Students expressed concerns about privacy and use of data.

**Aspirations**: Finally, although the findings from this study provide a relatively balanced view on student perspectives, it also raises some concerns about the unusually high expectations of some students for the role GenAI can play in their education. This is especially reflected in their views on GenAI's role in personalizing learning for them. They expect that the technology will be able to assess their knowledge and learning ability and tailor learning for them. Students expressed that this is important given that faculty are teaching many students and do not have the time to support individual learning. GenAI, for them, can fill the space in the middle and mitigate any lack of teaching through tutoring.

**Implications**: This study has some important implications for educators and educational institutions. The use of GenAI, at least at the experimental level, is high among students; therefore, there needs to be some consensus around its use within universities. Furthermore, faculty and instructors need to develop expertise and experiment with these systems to assess their usefulness, capabilities, and boundaries. This is important for teaching and learning and preparing students for the workforce. There must be an assessment of the optimum and useful use of these technologies, how students should be guided and scaffolded, and at what point their use should be prohibited.

There are also important implications for the design of assessments. It is evident that GenAI applications, especially ChatGPT and Co-Pilot, are being used by students at a relatively larger scale to complete assignments and even to plagiarize. As the applications improve, this trend is likely to continue, and therefore, there is a need to re-design assessments or incorporate reflective thinking related to GenAI use as part of the assignments [19]. There is also a need to revisit the issue of cheating and plagiarism within engineering education, given the rise of GenAI applications [20].

**Limitations and Future Work**: This is a small study with a limited sample size of 37 students; therefore, the findings have limited generalizability. Yet, compared to prior work on student perceptions of GenAI, this work adds to the literature by highlighting the value tensions and misconceptions related to GenAI among technology students. These themes, we believe, are highly relevant and of interest more generally. Finally, this study only tangentially focuses on ethical issues of using GenAI and future work needs to address this concern more centrally [21], [22].

## VI. Conclusion

The use of GenAI in education is an important consideration and a topic much in flux, given the dynamic nature of the technology and the lack of consensus on its application for teaching and learning. Therefore, it is important to understand what different stakeholders perceive as potential benefits and concerns related to it. It is especially important to understand students' viewpoints as they are a category of early users. In this paper, we present findings that show that although many students use it, they have misconceptions about how the technology works and what it is capable of and, thus, have high expectations and aspirations that are not in tune with reality. They also have value tensions around its use, particularly in relation to the ease of plagiarism and the lack of learning when using it to make things easy for themselves. Finally, many students expressed aspirations about how GenAI can help them learn that are not in line with what the technology is capable of. Findings from this work can help institutions and instructors design better ways of incorporating GenAI in teaching.

ACKNOWLEDGMENT

We thank the study participants. This work partly supported by U.S. NSF Award U.S. NSF Awards 2319137, 1954556 and USDA/NIFA Award 2021-67021-35329. Any opinions, findings, and conclusions or recommendations expressed in this material are those of the authors and do not necessarily reflect the views of the funding agencies.

REFERENCES

[1] N. McDonald, A. Johri, A. Ali, and A. Hingle, "Generative artificial intelligence in higher education: Evidence from an analysis of institutional policies and guidelines," *arXiv preprint arXiv:2402.01659*, 2024.
[2] A. Johri, A. S. Katz, J. Qadir, and A. Hingle, "Generative artificial intelligence and engineering education." *Journal of Engineering Education*, vol. 112, no. 3, 2023.
[3] N. Ranade, M. Saravia, and A. Johri, "Using rhetorical strategies to design prompts: a human-in-the-loop approach to make ai useful," *AI & SOCIETY*, pp. 1–22, 2024.
[4] O. Almatrafi, A. Johri, and H. Lee, "A systematic review of ai literacy conceptualization, constructs, and implementation and assessment efforts (2019-2023)," *Computers and Education Open*, p. 100173, 2024.
[5] C. K. Lo, "What Is the Impact of ChatGPT on Education? A Rapid Review of the Literature," *Education Sciences*, vol. 13, no. 4, p. 410, Apr. 2023. [Online]. Available: https://www.mdpi.com/2227-7102/13/4/410
[6] S. Grinschgl, F. Papenmeier, and H. S. Meyerhoff, "Consequences of cognitive offloading: Boosting performance but diminishing memory," *Quarterly Journal of Experimental Psychology*, vol. 74, no. 9, pp. 1477–1496, 2021.
[7] C. K. Y. Chan, "A comprehensive ai policy education framework for university teaching and learning," *International journal of educational technology in higher education*, vol. 20, no. 1, p. 38, 2023.
[8] J. B. Biggs, *Teaching for quality learning at university: What the student does*. Open university press, 1999.
[9] C. K. Y. Chan and W. Hu, "Students' voices on generative AI: perceptions, benefits, and challenges in higher education," *Int J Educ Technol High Educ*, vol. 20, no. 1, p. 43, Jul. 2023.
[10] H. Singh, M.-H. Tayarani-Najaran, and M. Yaqoob, "Exploring Computer Science Students' Perception of ChatGPT in Higher Education: A Descriptive and Correlation Study," *Education Sciences*, vol. 13, no. 9, p. 924, Sep. 2023. [Online]. Available: https://www.mdpi.com/2227-7102/13/9/924
[11] S. Ivanov, M. Soliman, A. Tuomi, N. A. Alkathiri, and A. N. Al-Alawi, "Drivers of generative AI adoption in higher education through the lens of the Theory of Planned Behaviour," *Technology in Society*, vol. 77, p. 102521, Jun. 2024. [Online]. Available: https://www.sciencedirect.com/science/article/pii/S0160791X24000691
[12] A. Yusuf, N. Pervin, and M. Román-González, "Generative ai and the future of higher education: a threat to academic integrity or reformation? evidence from multicultural perspectives," *International Journal of Educational Technology in Higher Education*, vol. 21, no. 1, p. 21, 2024.
[13] C. K. Y. Chan and K. K. W. Lee, "The AI generation gap: Are Gen Z students more interested in adopting generative AI such as ChatGPT in teaching and learning than their Gen X and millennial generation teachers?" *Smart Learn. Environ.*, vol. 10, no. 1, p. 60, Nov. 2023. [Online]. Available: https://slejournal.springeropen.com/articles/10.1186/s40561-023-00269-3
[14] A. Johri and A. Hingle, "Students' technological ambivalence toward online proctoring and the need for responsible use of educational technologies," *Journal of Engineering Education*, vol. 112, no. 1, pp. 221–242, 2023.
[15] A. Czeskis, I. Dermendjieva, H. Yapit, A. Borning, B. Friedman, B. Gill, and T. Kohno, "Parenting from the pocket: Value tensions and technical directions for secure and private parent-teen mobile safety," in *Proceedings of the sixth symposium on usable privacy and security*, 2010, pp. 1–15.
[16] E. Huhtamo *et al.*, "The self-driving car: A media machine for posthumans?" *Artnodes*, no. 26, pp. 1–14, 2020.

[17] Y. Cai, P. Jing, B. Wang, C. Jiang, and Y. Wang, "How does "over-hype" lead to public misconceptions about autonomous vehicles? a new insight applying causal inference," *Transportation research part A: policy and practice*, vol. 175, p. 103757, 2023.

[18] M. Vaismoradi, J. Jones, H. Turunen, and S. Snelgrove, "Theme development in qualitative content analysis and thematic analysis," *Journal of Nursing Education and Practice*, vol. 6, no. 5, 2016.

[19] B. M. Olds, B. M. Moskal, and R. L. Miller, "Assessment in engineering education: Evolution, approaches and future collaborations," *Journal of Engineering Education*, vol. 94, no. 1, pp. 13–25, 2005.

[20] D. D. Carpenter, T. S. Harding, C. J. Finelli, S. M. Montgomery, and H. J. Passow, "Engineering students' perceptions of and attitudes towards cheating," *Journal of Engineering Education*, vol. 95, no. 3, pp. 181–194, 2006.

[21] A. Johri, E. Lindsay, and J. Qadir, "Ethical concerns and responsible use of generative artificial intelligence in engineering education," in *Proceedings of the 51st SEFI Annual Conference*, 2023.

[22] E. D. Lindsay, M. Zhang, A. Johri, and J. Bjerva, "The responsible development of automated student feedback with generative ai," 2024. [Online]. Available: https://arxiv.org/abs/2308.15334